\documentstyle[epsfig]{mn2e}

\title[Quantifying Cosmic Superstructures]{Quantifying Cosmic Superstructures}
\author[J.\ M.\ Colberg]
       {J\"org M.\ Colberg$^{1}$\thanks{E-mail: astro@jmcolberg.com}\\
        $^1$ Carnegie Mellon University, Department of Physics,
             5000 Forbes Avenue, Pittsburgh PA 15213, USA}
\date{Accepted 2006 November 15. Received 2006 September 8; in original form 2006 May 4}

\pagerange{\pageref{firstpage}--\pageref{lastpage}}
\pubyear{2006}

\begin{document}

\maketitle

\label{firstpage}

\begin{abstract}
The Large Scale Structure (LSS) found in galaxy redshift surveys and in 
computer simulations of cosmic structure formation shows a very complex  
network of galaxy clusters, filaments, and sheets around large voids. 
Here, we introduce a new algorithm, based on a Minimal Spanning Tree, 
to find basic structural elements of this network and their properties.
We demonstrate how the algorithm works using simple test cases and then apply
it to haloes from the Millennium Run simulation (Springel et al. 2005). 
We show that about 70\% of the total halo mass is contained in a structure 
composed of more than 74,000 individual elements, the vast majority of which 
are filamentary, with lengths of up to 15\,$h^{-1}$\,Mpc preferred. 
Spatially more extended structures do exist, as do examples of what appear 
to be sheet--like configurations of matter. What is more, LSS appears to 
be composed of a fixed set of basic building blocks. The LSS formed by 
mass selected subsamples of haloes shows a clear correlation between the 
threshold mass and the mean extent of major branches, with cluster--size 
haloes forming structures whose branches can extend to almost 
200\,$h^{-1}$\,Mpc -- the backbone of LSS to which smaller branches 
consisting of smaller haloes are attached.
\end{abstract}


\section{Introduction}

Galaxy redshift surveys such as the Sloan Digital Sky Survey (York 
et al. 2000) and the 2dFGRS (Colless et al. 2001) show that galaxies are 
spread out in a fairly complicated way, over the so--called Cosmic Web. 
This network consists of the largest non--linear structures in the Universe,
galaxy clusters, which are interconnected through filaments and sheets. 
Embedded in this network are vast regions that contain almost no galaxies, 
so--called voids. 

N--body simulations of cosmic structure formation (for example Springel et al. 
2005) have been able to reproduce the {\it appearance} of the matter 
distribution very well; and the interplay between theoretical simulations and 
observational results has contributed to a large extent to our knowledge of 
the properties of the $\Lambda$CDM concordance model. 

Describing the network and comparing observations and simulations is no 
easy task. The two--point and, to a much lesser degree, higher--order 
correlation functions (e.g. Peebles \& Groth 1975; Peebles 1980; 
Peacock 1999) have been used most frequently. Other tools include Minimal 
Spanning Trees (see e.g. Barrow et al. 1985; Bhavsar \& Splinter 1996; 
Krzewina \& Saslaw 1996), the genus statistics (Gott et al.\ 1986; 
Springel et al. 1998), shape statistics (see e.g. Babul \& Starkman 
1992; Luo \& Vishniac 1995; Luo et al. 1996), and Minkowski functionals 
(Mecke et al. 1994; for a very detailed review see Sheth \& Sahni 2005 
and references therein).

Opinions about the nature and elements of Large--Scale Structure (LSS) 
differ to some extent. Are filaments or sheets the dominant structural 
elements? In a recent study, Colberg et al. (2005) investigated the 
configurations of matter between neighbouring clusters in an N--body 
simulation. They found a very strong preference for filaments over
sheets for those pairs of clusters whose connection was not cutting 
through a void. The existence of both filaments and sheets is very 
encouraging. This is because the visual impression from large redshift 
surveys indicates a filamentary network that includes very prominent 
sheets such as the ``Sloan Great Wall'' (Vogeley et al. 2004). Colberg 
et al. (2005) also reported on sizes of inter--cluster filaments, incl. 
averaged density profiles. As it turns out, the averaged densities agree 
very well with predictions of an analytical model by Shen et al. (2005).

However, the method employed by Colberg et al. (2005) has its problems.
First, given that they investigated inter--cluster matter configurations
by eye, the method is simply not feasible for larger data sets than
those used in their study\footnote{Visually inspecting thousands of
such inter--cluster matter configurations took about two weeks.}. Second, 
searching for structure elements between neighbouring clusters places a 
restriction on LSS. While massive clusters do indeed appear to lie at the 
intersections of filaments, it is not inconceivable that, for example, 
filaments might actually meet without a cluster being present.

With these restrictions in mind, the goal of this work is to devise an
algorithm that can classify LSS and that can decompose LSS into individual 
elements without any assumptions {\it a priori}.

\begin{figure}
\includegraphics[width=85mm]{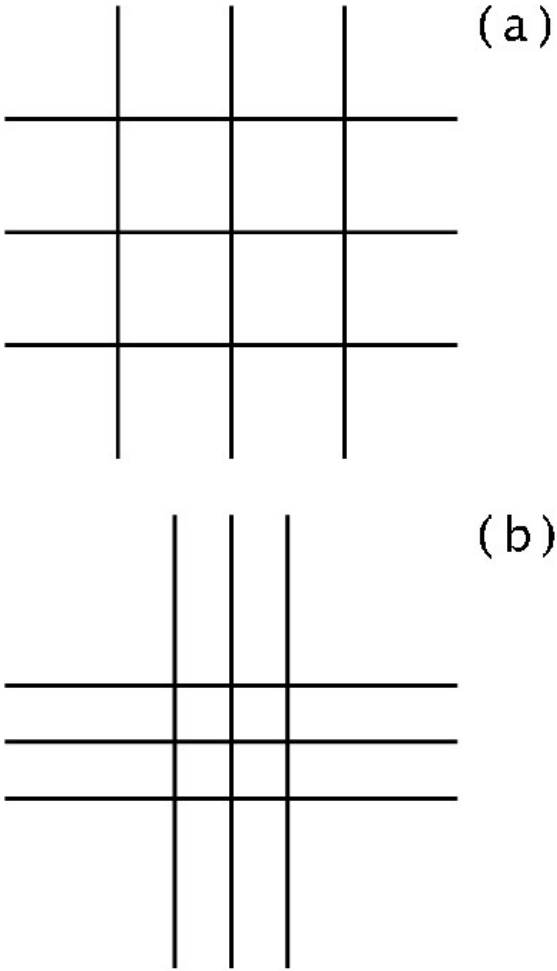}
\vspace{-1.0cm}
\caption{Two very simple model cases of a network of filaments. Both
         examples contain the same number of filaments, intersecting in
         a regular fashion, albeit with different separations from each
         other.}
\label{two_cases}
\end{figure}

As already mentioned above, there are other methods that have been
used frequently to describe LSS, with Minkowski functionals (Mecke 
et al. 1994) being the most common one. One immediate problem with this 
approach, however, is that the four Minkowski functionals are a 
topological measure of structure and thus do not directly yield sizes 
of objects. Because of this, so--called shapefinders derived from them 
are frequently used (Sahni et al. 1998). For simple toy models, these 
quantities deliver very clear and easily testable results. However, 
for cosmological data the situation usually gets a bit murky. With the 
raw data usually being smoothed on scales of 5.0\,$h^{-1}$\,Mpc or 
more\footnote{We express the Hubble constant in units of $H_0 = 
100\,h$\,km\,sec$^{-1}$\,Mpc$^{-1}$.} and {\it averages} over the whole 
volume, Minkowski Functionals and related shapefinders are not able to 
deliver the detailed kind of information that we are interested in. 

What is more, Minkowski Functionals are actually unsuited for What we 
are after here. Take the two very simple examples shown in 
Figure~\ref{two_cases}. Both cases contain six model filaments (that
are much longer than wide), which form a simple network. They both
yield the exact same Minkowski functionals, whereas they {\it look} 
different: In case a, each intersection is connected to four identical
fragments, whereas in case b, each intersection is connected to either 
two short and two long or three short and one long or four short 
fragments. If we take Figure~\ref{two_cases} as a
very simplistic model case for the inter--cluster filaments studied in 
Colberg et al. (2005), it becomes obvious that Minkowski functionals 
lack predictive power to tell the difference between the two cases.

The aim of this work is to locate structure elements of LSS and to
describe their properties. This is done by running a group finder on 
haloes from a very large and very detailed N--body simulation, and 
by then directly determine the actual geometrical structure of those 
objects. The algorithm consists of the application of well--known 
techniques such as a friends--of--friend group finder and a Minimal 
Spanning Tree (MST) plus a new way to categorize the MST and to 
extract its branches and their properties. We will then apply the 
algorithm to the simple test cases shown above and to the full set 
of haloes from the Millennium Run simulation (Springel et al. 2005). 

This paper is organized as follows. In the following Section
(\ref{simulation}), we first briefly introduce the simulation data used
in this work. In Section~\ref{finding}, we discuss the algorithm to
find cosmic structures, which consists of  using a group finder 
(Section~\ref{groupsfinding}), a Minimal Spanning Tree (Section~\ref{mst}),
and a new algorithm to classify the elements of the latter 
(Section~\ref{locating}). In Section~\ref{results}, we present
results of the study, namely numbers of structural elements 
(Section~\ref{numbers}), their shapes (Section~\ref{shapes}), and
their spatial extents (Section~\ref{extents}). Section~\ref{massdep} 
contains a study of the mass dependence, and in Section~\ref{sampling},
we study the influence of sampling. Finally, in Section~\ref{discussion} 
we discuss the findings of the study. 

\section{The Simulation} \label{simulation}

\subsection{Simulation Details}

We make use of the $z=0$ halo catalogue obtained from the Millennium Run,
a very large dark--matter--only N--body simulation of the concordance 
$\Lambda$CDM cosmology with 2160$^3$ particles in a (periodic) box of 
size 500\,$h^{-1}$\,Mpc in each dimension. The cosmological parameters 
are total matter density $\Omega_m = 0.25$, dark energy/cosmological 
constant $\Omega_\Lambda=0.75$, Hubble constant $h = 0.73$, and the 
normalization of the power spectrum $\sigma_8 = 0.9$. With these parameters, 
each individual dark matter particle has a mass of $8.6 \cdot 
10^8\,h^{-1}$\,M$_{\odot}$. See Springel et al. (2005) for more details 
of the simulation.

\subsection{Haloes and subhaloes}

The simulation code was designed to produce friends--of--friends (FOF)
group catalogues on the fly and to output these along with particle
data at each output time. FOF groups are sets of particles in which
any pair's separation does not exceed some fraction $b$ of the mean
inter--particle separation (Davis et al. 1985, but also see the
earlier application in Turner \& Gott 1976). With a choice of $b=0.2$,
the groups have mean overdensities of about 200, which roughly corresponds
to that expected for a virialized group. There are $17.7 \cdot 10^{6}$ 
FOF groups in the simulation volume, each with 20 particles or more.

Given the very high mass resolution of the Millennium Simulation, the
haloes contain substructures, which, however, are not resolved with the
FOF group finder. To find the substructures, an improved and extended
version of SUBFIND (Springel et al. 2001) was applied. That way, a total
sample of $18.2 \cdot 10^{6}$ subhaloes was found. Here, a halo is counted
as a single subhalo if it does not contain any identifiable substructure.
The largest halo contains 2328 subhaloes. For more details on the group
finding again see Springel et al. (2005).

\section{Finding Large--Scale Structures} \label{finding}

\subsection{Overview: Locating Cosmological Superstructures} \label{locatingstructures}

The procedure described in the following aims at locating structures in the 
Millennium Run simulation in such a way that the structure elements agree 
with those found by eye (for interesting discussions of this see Barrow \& 
Bhavsar 1987 and Bhavsar \& Ling 1988b). With such a seemingly vague {\it 
ansatz} there are no obvious physical criteria on how to proceed. It would 
be tempting to adopt simple toy models and to base the procedure on those. 
However, cosmological structures appear to be far more complicated than a 
simple mix of simplified cylinders (filaments) and sheets (walls); and we 
want to avoid biasing the study by relying on too simple models.


What the eye (or rather the brain) does is to take the massive clusters as
nodes of the network, and it classifies the chains of haloes in between 
them into different categories, depending on where they lie and how close
they are to their neighbours. Here, we will mimick this kind of classification
as follows:
\begin{enumerate}
  \item Groups of haloes are found using a standard fof group finder for the 
        full halo sample. Given the fact that the linking length is a free 
        parameter, this procedure will be repeated with different choices for 
        the linking length. We focus our attention to the parameter range 
        where the vast majority of haloes ends up in one big group.
  \item The haloes in the largest group are assigned to a three--dimensional
        grid. The size of its cells are chosen on the basis of the smallest 
        scales to be resolved. We here pick 2\,$h^{-1}$\,Mpc in each dimension, 
        since we are only interested in the large--scale distribution of 
        matter. Note that this step reduces the computational complexity of 
        the following steps by decreasing the number of the data points. 
  \item We construct the Minimal Spanning Tree (MST) for those cells that 
        contain at least one halo. The technical details of a MST will be 
        discussed further below; briefly summarized a MST is a data structure 
        that contains nodes (in our case grid cells containing haloes) and 
        connections (so--called edges) between them, with the sum of the 
        lengths of the edges being minimal.
  \item We classify the nodes of the MST according to their position inside
        the MST as major and minor nodes (again, details below). The former 
        are nodes such as those formed by massive galaxy clusters. These lie 
        at the intersections of filaments and sheets, where the latter can be
        found.
  \item Having classified the MST that way, we end up with a hierarchical
        data structure of branches and subbranches, from which we can
        determine the sizes and shapes of structure elements quite easily.
\end{enumerate} 

In the following Sections, we will discuss the individual steps in more
detail.

\subsection{Group Finding} \label{groupsfinding}

The first step of the procedure to find LSS elements is to run a standard 
FOF group finder on the halo catalogue. For the linking 
length\footnote{Note that for large haloes we use their subhaloes for the 
group finding. Since the virial radii pf cluster--size haloes are larger 
than the linking length, using only haloes would exclude those haloes from 
the search. Using the subhaloes solves this problem.}, we adopt a 
variety of fractions of the mean interhalo separation $l=2.01\,h^{-1}$\,Mpc, 
ranging from values between 0.5 and 0.6. Our choice of linking length is 
motivated by the following. 

Between values of 0.5 and 0.6 times the mean inter--halo separation, we 
witness the onset of percolation: Initially, the largest object contains 
only a couple of percent of the total (halo) mass. The fraction of mass 
in the largest objects then grows very rapidly to finally reach 70\% of 
the total mass (see Table~\ref{table_results}). For point distributions, 
this process has not been studied in much detail since Dekel \& West (1985) 
concluded that the strong dependence of the onset of percolation on the mean 
density of the sample ruled out its use as a tool to study LSS. For
earlier applications of percolation to study cosmic structures see
the seminal papers by Bhavsar \& Barrow (1983), Einasto et al. (1983), 
and Shandarin (1983), also Klypin \& Shandarin (1993).

For the present study, the exact choice of linking length does not matter
as long as we stay in the regime where the largest object contains
a significant amount of LSS. We will later study systematic effects such
as different linking lengths or different halo sample sizes.


The upper--left panel of Figure~\ref{fig:slicemosaic} shows a slice of
thickness 15\,$h^{-1}$\,Mpc through the simulation volume, with the sizes 
of the symbols reflecting the halo masses. The symbols 
show haloes that are part of the largest object at 60\% of the mean 
interhalo separation. Note that because this image shows a thin slice 
through the volume, the largest object appears to be broken up into 
separate pieces. With this choice of linking length, the largest object 
contains 69.4\% of the total mass. 

Since we are interested in the sizes and shapes of structures on large scales 
we bin the haloes of each group onto a three--dimensional grid with a cell 
size of (2.0\,$h^{-1}$\,Mpc$)^3$. We represent each grid cell that includes 
at least one halo by a single particle whose position corresponds to the 
center of the cell.

\subsection{Minimal Spanning Trees} \label{mst}

Twenty years ago, Barrow et al. (1985) introduced the Minimal Spanning Tree
(MST) into the astronomical context. Initially thought to be an interesting 
tool, MSTs have not been used much over the course of the past ten years (see, 
for example, Bhavsar \& Splinter 1996, Krzewina \& Saslaw 1996, Doroshkevich et 
al. 2001, Demianski \& Doroshkevich 2004), probably because of appeal of Minkowski 
Functionals and derived quantities. Bhavsar \& Ling (1988) used MSTs to
identify filaments in the CfA catalog, providing the first statistical
evidence for the existence of filaments.

A Minimal Spanning Tree is a technique from graph theory (for the following brief 
review also see Barrow et al. 1985 and references therein). A graph is a collection 
of {\it nodes} (in astronomical cases typically galaxies), {\it edges}
(straight connections between nodes), and edge lengths. A {\it path} is a
sequence of edges that join nodes. A closed path is called a {\it circuit}; and
a graph is called {\it connected} if all points are included in a path. A 
connected graph without any circuits is called a {\it tree}. If the tree of a 
connected graph contains all nodes, the tree is called a {\it spanning tree}. 
A tree's length is defined as the sum of the edge lengths. The minimal 
spanning tree then is the spanning tree whose length is smallest. If no two edge 
lengths are equal the MST will be unique\footnote{Given that we are working
with grid--based data, the MST in this work is not unique. We ran some tests
with different MST's that all represented the same data to find no differences
in the properties of the structure elements constructued by our algorithm.}. 
For more details on MSTs please c.f.\ Barrow et al. 1985 and references therein.


There are a few reasons why we are using the MST of the groups of haloes. Most 
importantly, by construction a MST does not contain any loops. This property 
is of crucial importance for the algorithm to locate branches of the tree,
as will become clear in the following.

\subsection{Locating Tree Elements} \label{locating}

Once we have computed the MST we can apply our algorithm to find the branches. 
In the past (see, for example, Barrow et al. 1985), MSTs were typically pruned 
and broken apart. Smaller branches were cut off, and the MST was broken into 
pieces by removing the links between nodes that are connected by an edge of 
some given length. We will refrain from doing this here. 

The idea behind the algorithm is very simple: The algorithm is designed to 
locate branches in such a way that smaller branches are part of larger ones 
and the whole tree is divided into branches naturally. 

A detailed discussion of how the classification algorithm works can be found
in the Appendix. Briefly summarized, the algorithm visits each node in the
MST, and it determines the relation of the node to its neighbouring nodes.
Depending on their numbers of neighbours, nodes are grouped into branches,
with smaller branches becoming members of larger ones.

\begin{figure}
\includegraphics[width=85mm]{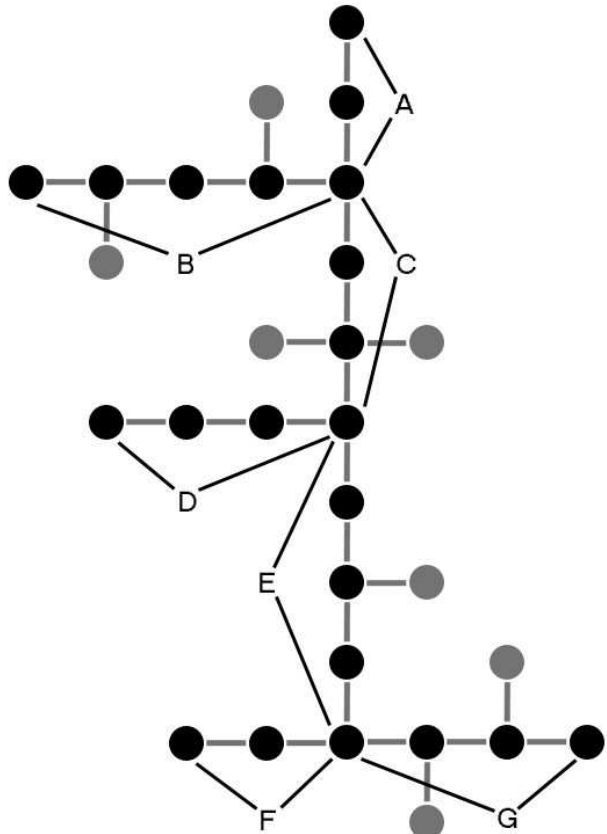}
\vspace{-1.0cm}
\caption{Major branches for a simple tree. In this example, a major branch has
         to have at least length 2. Nodes in major branches are shown in
         black, the others in grey. There are seven major branches (labeled A
         to G). Note that in this very simple example, all major branches are
         straight.} 
\label{major_branches}
\end{figure}

After the classification of nodes and the creation of hierarchical sets of
branches, the code categorizes the structure by dividing it into {\it major} 
and {\it minor} branches. Major branches are those that are longer than a 
given (arbitrary) length $l$. This procedure is the equivalent of pruning an 
MST. Figure~\ref{major_branches} shows a very simple example.

For the analysis of the halo groups, we choose $l = 10\,h^{-1}$\,Mpc or
five mesh cells. But note that with this procedure major branches can be
shorter than $10\,h^{-1}$\,Mpc. This will be the case for those branches that 
lie inside the structure and have no loose end. 

\subsection{Simple Test Cases} \label{testcases}

\begin{figure}
\includegraphics[width=85mm]{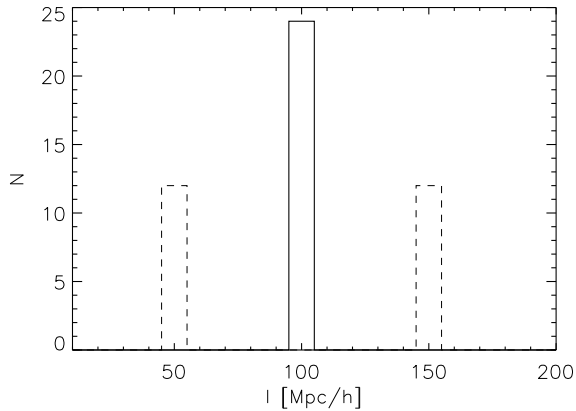}
\vspace{-0.7cm}
\caption{Size distributions of the major branches for the two cases shown in
         Figure~\ref{two_cases}, with the solid (dashed) histogram showing
         case $a$ ($b$). See text for details.} 
\label{grid_test}
\end{figure}

\begin{figure}
\begin{center}
\includegraphics[width=75mm]{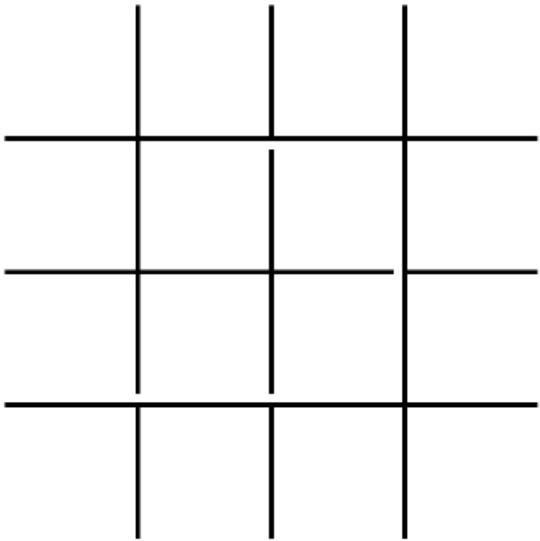}
\end{center}
\caption{MST constructed by our software for the first of the two simple
         cases shown in Figure~\ref{two_cases}} 
\label{MST_Case_A}
\end{figure}

As a proof of concept, we first apply the structure--finding algorithm to the
two cases shown in Figure~\ref{two_cases}. Case $a$ and $b$ consist of six 
long intersecting lines each, with nine vertices. These intersections divide 
the lines into segments, and it is these segments that the algorithm finds.

For both cases, we created a set of 120,000 model particles, which we 
distribute randomly, following the spatial patterns shown in Figure~\ref{two_cases}.
For case $a$, the spacings of the model filaments were chosen to be 
100\,$h^{-1}$\,Mpc. For case $b$, the filaments were spaced 50\,$h^{-1}$\,Mpc 
apart. Thus, in case $a$, the segments all have the same length 
(100\,$h^{-1}$\,Mpc), whereas in case $b$, half the segments have a length of 
50\,$h^{-1}$\,Mpc and half are 150\,$h^{-1}$\,Mpc long. 

Figure~\ref{grid_test} plots the size distributions of the major branches 
measured for the model particle distributions for case $a$ (solid histogram) 
and $b$ (dashed histogram). As can be seen from the Figure, our algorithm 
identifies the different segments as major branches and {\it exactly}
reproduces their size distributions. It needs to be stressed that the
two cases as shown in Figure~\ref{two_cases} are {\it not}
MST's\footnote{Remember that an MST does not contain any closed loops!}. 
As an example, Figure~\ref{MST_Case_A} shows the MST constructed by our code 
for case (a). The code placed the small gaps needed to convert the graphs
from Figure~\ref{two_cases} into MST always right next to the vertices,
which led to the exact reproduction of the size distributions. In principle,
it is possible that the code will break a connection between vertices 
somewhere in the middle, thus breaking a structure element into two
pieces. It is impossible for us to exactly quantify to what extent this is
actually happening in the vastly more complex cases from the simulation
discussed below. However, it is not likely that the misidentification of 
structure elements as broken pieces will lead to a significant distortion
of our results, since the structure of the graphs used as simple test 
cases from Figure~\ref{two_cases} is {\it much} simpler than those of 
actual simulation data. There, unlike in the toy models, loop--like 
structures formed by haloes tend to contain a very large number of 
individual structure elements. Breaking some of them into two pieces
will thus introduce only a very small error.

In addition to the two simple cases shown in Figure~\ref{two_cases}, we tested
the algorithm on a number of more complex, fully three--dimensional geometric
configurations of model particles. In each of those cases, the algorithm was 
able to determine the correct numbers, shapes, and sizes of the individual 
structure elements. 

We now apply the structure finding algorithm to the Millennium Run halo
sample and study the results.

\begin{table}
  \begin{center}
  \begin{tabular}{lrr}
    \hline
     $b$ & $f_m$ & $n$ \\
    \hline
     0.50 &  2.1\% &  1191 \\
     0.51 &  6.9\% &  4175 \\
     0.52 & 14.2\% &  9543 \\
     0.53 & 29.9\% & 21288 \\
     0.54 & 38.5\% & 29485 \\
     0.55 & 47.4\% & 38591 \\
     0.56 & 53.5\% & 46059 \\
     0.57 & 58.2\% & 53422 \\
     0.58 & 62.5\% & 60584 \\
     0.59 & 65.9\% & 67282 \\
     0.60 & 69.4\% & 74086 \\
    \hline 
  \end{tabular}
  \end{center}
  \caption{Sizes of the largest groups of haloes as a function of the fraction
           $b$ of the mean inter--halo separation. $f_m$ is the fraction of
           total halo mass that is contained in the largest group. $n$ denotes
           the number of major and minor branches in the group (see text for details).}
  \label{table_results}
\end{table}

\begin{figure}
\includegraphics[width=85mm]{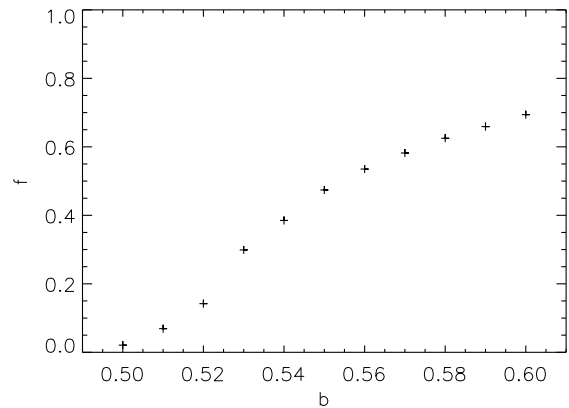}
\vspace{-0.7cm}
\caption{Fraction of halo mass contained in the largest object as a function
         of the linking parameter $b$.} 
\label{perc}
\end{figure}

\begin{figure}
\includegraphics[width=85mm]{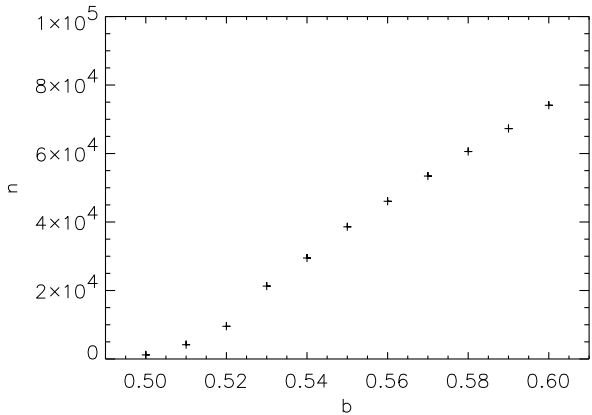}
\vspace{-0.7cm}
\caption{Number of major branches in the largest object as a function
         of the linking parameter $b$.} 
\label{n_branches}
\end{figure}

\begin{figure}
\includegraphics[width=85mm]{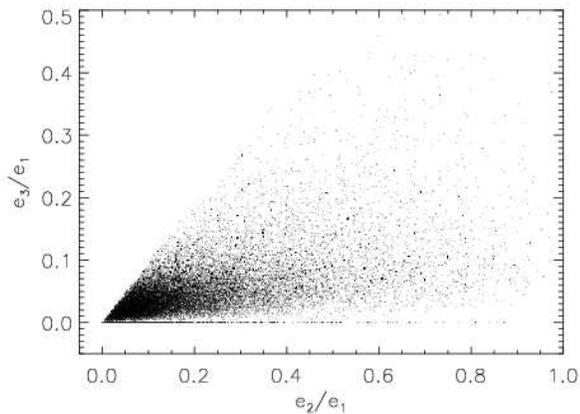}
\vspace{-0.7cm}
\caption{Shape distributions of major branches ($b = 0.6$). The plot shows the ratio
         $e_2/e_1$ versus the ratio $e_3/e_1$.} 
\label{shapes_total}
\end{figure}

\begin{figure}
\includegraphics[width=85mm]{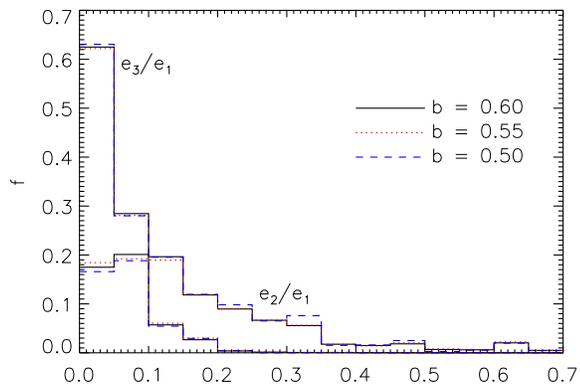}
\vspace{-0.7cm}
\caption{Fractional distributions of $e_2/e_1$ and $e_3/e_1$ for major
         branches ($b = 0.6$). Shown are distributions for $b=0.60$ (solid line),
         $b=0.55$ (dotted line), and $b=0.50$ (dashed line).} 
\label{shapes_histo}
\end{figure}

\section{Results obtained for Millennium Run halo samples} \label{results} 

Table~\ref{table_results} summarizes the fraction $f_m$ of the total halo mass
contained in the largest object and the number of major branches as a 
function of the linking parameter $b$. The effect of percolation is 
shown in Figure~\ref{perc} (c.f.\ Section~\ref{groupsfinding}). $f_m$ 
rises from just a couple of percents at $b = 0.5$ to almost 70\% at 
$b = 0.6$. We also find that at any $b$ the largest object always is 
much larger than the second largest or any other object in the volume. 

\subsection{Numbers of structure elements} \label{numbers}

As we increase $b$, so does the number of haloes in the largest object. 
Figure~\ref{n_branches} shows how this increase in size translates into 
the number $n$ of major branches. At the largest $b$, there are more than 
74,000 major structure elements present. This finding impressively confirms 
the immense complexity of LSS.

At $b = 0.53$ and above the relation between $f_m$ and $n$ is almost linear. 
The growth of the largest object around the percolation threshold can be 
understood as follows. As as the linking length increases, a larger and
larger number of formerly disjoint objects that gets interconnected. The 
linear relationship between $b$ and $n$ in Figure~\ref{n_branches} clearly
reflects this process. We will study next whether the properties of these 
major branches change with $b$. 

\subsection{Shapes of major branches} \label{shapes}

To study the shapes of the branches\footnote{In the following, branches
will stand for major branches.} for we compute the quantity
\begin{equation}
I_{ij} = \sum\,x_{i}x_{j}
\end{equation}
for the centers of their cells, where the sum is over all cell centers. 
The normalized eigenvectors of $I$ correspond
to the unit vectors of the best--fit ellipse of the branch, and there are
three eigenvalues, sorted such that $e_1 > e_2 > e_3$. This procedure
is commonly used for the shapes of haloes (see, for example, Hopkins et al.
2005 and references therein). Note that because of the relatively small
number of cells per node\footnote{For example, for the $b = 0.6$ sample, the
largest major branch contains only 162 cells. Typically, $I$ is computed for
haloes, which -- depending on the simulation details -- contain many thousands
of particles.} $I$ provides only a fairly 
crude measure of the shapes of those branches. We will not be able to
determine more than whether branches are roughly filamentary, sheet--like, 
or elliptical. 

Figure~\ref{shapes_total} shows $e_2/e_1$ versus $e_3/e_1$ for the largest
sample ($b = 0.6$). Because $e_1 > e_2 > e_3$ the area with $e_3/e_1 >
e_2/e_1$ is empty. Note the different scales of the $x$ and $y$--axis. 
The vast majority of major branches has $e_3/e_1 < 0.1$, which means that 
they are planar or close to planar. Because of the nature of the branches 
some are even completely straight.

In Figure~\ref{shapes_histo}, we plot the fractional distributions of $e_2/e_1$ 
and $e_3/e_1$ for the three different halo groups $b=0.60$ (solid line),
$b=0.55$ (dotted line), and $b=0.50$ (dashed line). As already seen in
Figure~\ref{shapes_total}, just a few percent of branches have $e_3/e_1 >
0.1$. The distribution of $e_2/e_1$ is broader. However, only a few percent 
of the branches have $e_2/e_1 > 0.35$. $e_3/e_1 \approx 0.1$ and
$e_2/e_1 \leq 0.35$ corresponds to straight or slightly curved filaments. 

In addition to comparing the fractional distributions of $e_2/e_1$ and
$e_3/e_1$, Figure~\ref{shapes_histo} also compares halo groups for different
linking lengths $b$. There are no discernable differences between the samples.
The mean values of these quantities for the different samples
$\overline{e_2/e_1} = 0.1716$, 0.1734, and 0.1736 and $\overline{e_3/e_1} = 
0.0432$, 0.0437, and 0.0428 for $b = 0.60$, 0.55, and 0.50, respectively.

In a sense, the results obtained for the shapes of branches are not very 
surprising. Haloes that comprise the largest object at $b=0.50$ are also 
contained in the largest object at larger $b$. This statement is 
{\it not} trivial, though. It is possible to have a distribution where 
the haloes in the largest object at $b=0.50$, say, are not contained in 
the one at $b=0.55$. \footnote{This could happen if at $b=0.55$ a group of smaller 
objects from $b=0.50$ connect to form the largest object. However, this 
is not the case for our sample.}. 

Given the results so far, LSS thus can be understood as being composed of
a fixed set of building blocks, with different sizes and shapes, very
much like a set of cosmic Lego bricks. Depending on how one chooses the 
linking length, the largest object is constructed by picking the same 
pieces, albeit in different numbers. 

\begin{figure}
\includegraphics[width=85mm]{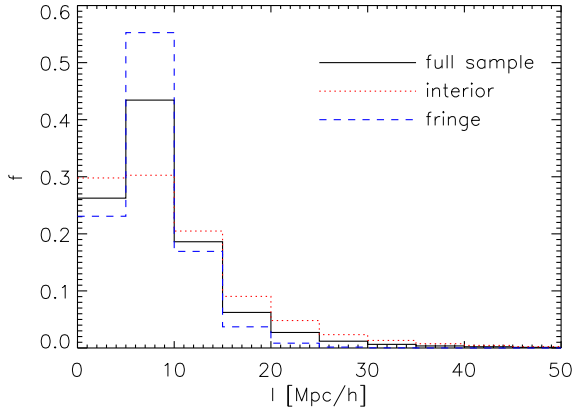}
\vspace{-0.7cm}
\caption{Distributions of extents of major branches ($b = 0.6$) for the whole structure
         (solid line) and for major branches that lie inside the structure 
         (dotted line) or that have a loose end (dashed line).} 
\label{histo_length}
\end{figure}

\begin{figure}
\includegraphics[width=85mm]{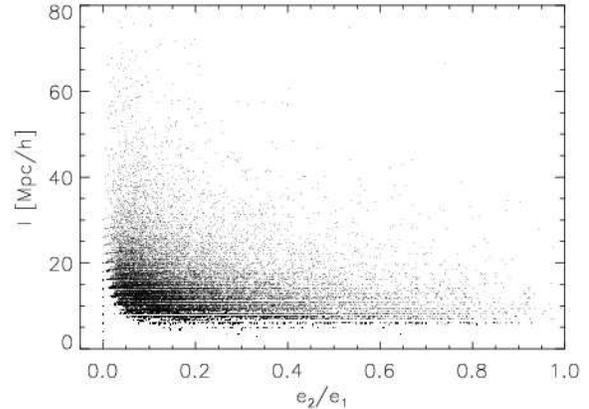}
\vspace{-0.7cm}
\caption{$e_2/e_1$ versus the extent for $b = 0.6$. The horizonal stripes are
         due to the coarse grid.} 
\label{shape_vs_length}
\end{figure}

\begin{figure*}
\begin{center}
\includegraphics[width=248mm]{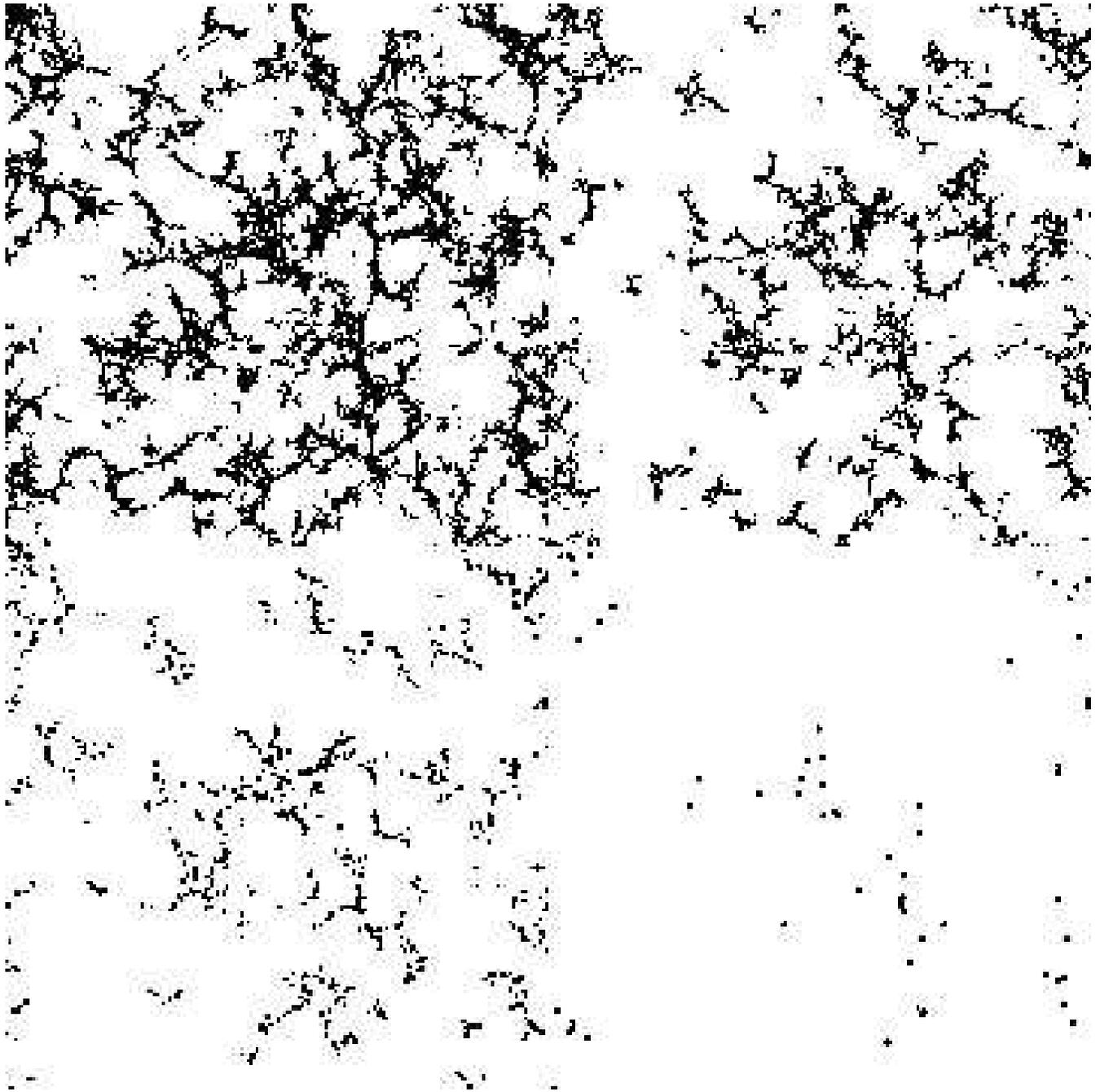}
\end{center}
\caption{Slices of thickness 15\,$h^{-1}$\,Mpc through the simulation volume. 
         Shown are only haloes that belong to the largest object, with 
         the size of the symbols reflecting the different halo masses. The 
         upper left panel uses all haloes. The upper right,
         lower left, and lower right panels show haloes belonging to the
         largest object for the halo samples with $m >
         10^{12}\,h^{-1}\mbox{M}_\odot$,  $m > 10^{13}\,h^{-1}\mbox{M}_\odot$,
         $m > 10^{14}\,h^{-1}\mbox{M}_\odot$, respectively. As discussed in
         the main text, the linking lengths used to construct the largest
         object for each halo sample lead to about 65\% of the halo sample
         mass contained in the largest object.}
\label{fig:slicemosaic}
\end{figure*}

\begin{figure}
\includegraphics[width=85mm]{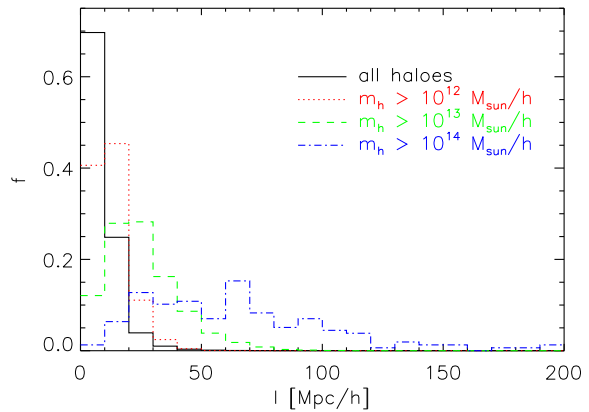}
\vspace{-0.7cm}
\caption{Distributions of extents of major branches for different halo
         samples. Shown are all haloes (solid line), haloes with
         $m > 10^{12}\,h^{-1}\mbox{M}_\odot$ (dotted line), haloes
         with $m > 10^{13}\,h^{-1}\mbox{M}_\odot$ (dashed line), and
         haloes with $m > 10^{14}\,h^{-1}\mbox{M}_\odot$ (dot--dashed
         line). In each case, the largest object whose major branches
         are used here contains about 65\% of the mass of the halo
         sample.} 
\label{extentshalosamples}
\end{figure}

\begin{table}
  \begin{center}
  \begin{tabular}{lrr}
    \hline
     sample & $N_{\mbox{\scriptsize major branches}}$ & $\bar{l}\,\,\,[h^{-1}$\,Mpc]\\
    \hline
     full sample & 71,064 & 9.1 \\
     $m > 10^{12}\,h^{-1}\mbox{M}_\odot$ & 15730 & 12.6 \\
     $m > 10^{13}\,h^{-1}\mbox{M}_\odot$ & 2431 & 25.7 \\
     $m > 10^{14}\,h^{-1}\mbox{M}_\odot$ & 193 & 64.4 \\
    \hline 
  \end{tabular}
  \end{center}
  \caption{Different halo samples used for the study of structural properties
           of Large--Scale Structure. For each sample, the largest object
           was studied at the linking length where it contained about
           65\% of the mass in the halo sample. $N_{\mbox{\scriptsize major branches}}$ 
           is the number of major branches, and $\bar{l}$ is the mean
           extent of the major branches (in $h^{-1}$\,Mpc).}
  \label{tablehalosamples}
\end{table}

\subsection{Spatial extents of major branches} \label{extents}

Figure~\ref{histo_length} shows the distributions of spatial extents of major 
branches. By spatial extent we here mean
\begin{equation}
l = \sqrt{(x_{max} - x_{min})^2 + (y_{max} - y_{min})^2 + (z_{max}-z_{min})^2}
\end{equation}
for each major branch, where ($x_{min}$, $y_{min}$, $z_{min}$) and ($x_{max}$,
$y_{max}$, $z_{max}$) are the minimum and maximum coordinates of the major
branch, respectively. For perfectly straight branches $l$ is the actual
length of the branch, for a branch that forms two sides of a rectangle, $l$
is the length of the diagonal, etc. 

We also divided the major branches into two categories. The first category (interior)
contains those branches that connect to two (or more) major branches at each
end. The second category (fringe) encompasses those that connect to two (or more) major 
branches only at one end. In other words, the latter branches have a loose
end. For example, in Figure~\ref{major_branches}, branches A, B, D, F, and G
belong to the second category, whereas only branches C and E are category one
branches. 

For the Millennium Run haloes, the numbers of branches in the two categories are
roughly equal. Figure~\ref{histo_length} shows the distributions of $l$ for
all (solid line), interior (dotted line), and fringe (dashed line) branches 
($b = 0.6$). Fringe branches appear to be shorter than interior ones. 

About two thirds of the major branches have extents of up to $l = 10\,h^{-1}$\,Mpc.
The other third extends to larger scales, with a very small number going 
beyond $l = 30\,h^{-1}$\,Mpc. 

In Figure~\ref{shape_vs_length} we plot $e_2/e_1$ versus $l$ for the $b = 0.6$ 
sample. Below $l = 20\,h^{-1}$\,Mpc the effect of the grid is clearly visible 
as horizontal stripes. There appears to be a tendency for shorter branches to 
have a broader distribution in $e_2/e_1$. This trend is almost entirely caused 
by how the $e$'s are computed. The shortest branches consist of only a few
cells, and thus their shapes have to be taken with a grain of salt. However, 
there is a general trend for more extended branches to be more 
filamentary\footnote{Remember that $e_3/e_1$ tends to be much smaller than 
$e_2/e_1$ -- see Figure~\ref{shapes_histo}.}. 

\subsection{Halo Mass Dependence} \label{massdep}

Having examined the properties of LSS formed by the full set of haloes,
we now turn our attention to subsamples. From the original halo set 
we construct three subsamples by requiring minimum masses of 
$m > 10^{12}\,h^{-1}\mbox{M}_\odot$,  $m > 10^{13}\,h^{-1}\mbox{M}_\odot$, 
and $m > 10^{14}\,h^{-1}\mbox{M}_\odot$. Very crudely, these masses 
correspond to those of late--type galaxies, groups of galaxies, and 
massive galaxy clusters, respectively. In the following, we will refer 
to these samples as m$_{12}$, m$_{13}$, and m$_{14}$.

In Figure~\ref{fig:slicemosaic}, we show a slice through the simulation 
volume, displaying only haloes that are part of the largest object. 
The upper left, upper right, lower left and lower right panel correspond 
to the full halo sample, and samples m$_{12}$, m$_{13}$, and m$_{14}$, 
respectively. In order to build the largest object we chose the 
individual linking lengths for the group finding in such a way that 
for each sample, the largest object contains about 65\% of the total 
mass in the halo sample. Table~\ref{tablehalosamples} summarizes the 
four samples; in the second column, the total number of haloes is quoted.

For each sample, we run the structure finder on the largest group and 
compute properties of the major branches. Columns three and four of 
Table~\ref{tablehalosamples} give the total number of major branches 
and their mean extent, respectively. As could be expected, the larger 
the minimum mass of the halo sample, the larger the mean extent of 
major branches. 

This effect is also visible in Figure~\ref{fig:slicemosaic}. More
massive haloes (sample m$_{14}$) can be predominantly found in overdense 
regions, and the bridges between them -- formed by less massive haloes -- 
are replaced with simple connections. In other words, many of the nodes 
visible in the upper panel of Figure~\ref{fig:slicemosaic} disappear as 
the mass threshold is increased, and for sample m$_{14}$ only the most 
massive nodes are left. 

Figure~\ref{extentshalosamples} provides a more detailed view of the 
extents of major branches. It shows histograms for the four samples. 
The extent of major branches in sample m$_{14}$ extends all the way 
up to a maximum value of 191.5\,$h^{-1}$\,Mpc. Very long, connected
chains of massive haloes are thus a feature of simulations of cosmic 
structure formation. This fact is particularly reassuring in the light 
of observations of very extended chains of galaxies and of cosmic 
superclusters and superstructures (see, for example, Bharadwaj
et al. 2004 or Gott et al. 2005). Images of the LSS in the simulation
(Figure~\ref{fig:slicemosaic}) clearly support the finding that there 
are coherent structures of this size. It is thus highly unlikely that
the sizes of the structure elements in the m$_{14}$ sample are due
to sparse sampling.

\begin{table}
  \begin{center}
  \begin{tabular}{lrr}
    \hline
     sample size & $f_m$ & $b$ \\
    \hline
     full sample & 0.649  &  0.600 \\
     50\% & 0.651 & 0.654 \\
     25\% & 0.650 & 0.699 \\
     10\% & 0.652 & 0.753 \\
    \hline 
  \end{tabular}
  \end{center}
  \caption{Different halo samples used to study sampling issues of structural properties
           of Large--Scale Structure. For each sample, we give the fraction of
           haloes used in the sample, the fraction of mass contained in the
           object studied, and the fraction $b$ of the mean inter--halo
           separation needed to produce that object.}
  \label{tablehalosampling}
\end{table}

\begin{figure}
\includegraphics[width=85mm]{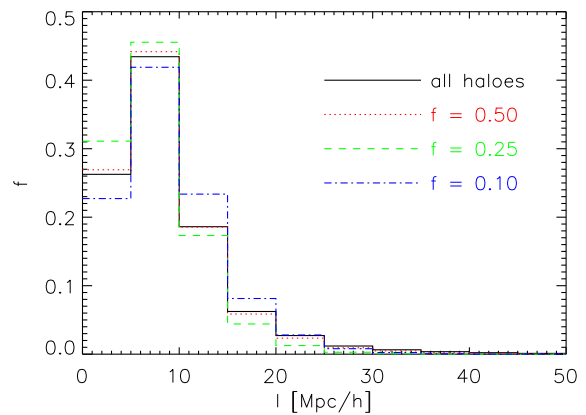}
\vspace{-0.7cm}
\caption{Distributions of extents of major branches for different halo
         samples. Shown are all haloes (solid line), and random subsamples 
         of different sizes: 50\% (dotted line), 25\% (dashed line), and
         10\% (dot--dashed line). In each case, the largest object whose 
         major branches are used here contains about 65\% of the mass of 
         the halo sample.} 
\label{extentshalosampling}
\end{figure}

\subsection{Sampling Issues} \label{sampling}

Having computed the extents of major branches both for the full halo
sample and for subsamples chosen by mass, it is important to test
how sample completeness affects the results. Before doing this
we need to address an important issues which might be raised as an
objection to the methods presented here.

Examining randomly selected subsamples of the haloes 
provides a test of the structure finding algorithm only if we make 
sure that similar objects are being compared. In the following, 
we will investigate objects that contain 65\% of the mass of each 
sample. As noted above, if the percolation of a set of points is
studied, there is a dependence of the percolation threshold on the
density of the point sample. What this means in a cosmological 
context is that if there are two samples of points with the same 
two--point correlation function but different densities, then the 
percolation threshold for the samples will be different. The sample 
with the lower density will have a higher percolation threshold. 
One can show that for very sparse samples, the percolation behaviour 
is that of a Poisson distribution (see Dekel \& West 1985). 

Table~\ref{tablehalosampling} summarizes the properties of
the different halo samples used for this test. As can be seen from
the first and last column, as the number density of the halo sample
decreases, the factor needed to scale the mean inter--halo separation
for the group finding increases.

In Figure~\ref{extentshalosampling}, we plot the extent distributions
of the major branches in the samples from Table~\ref{tablehalosampling}.
There is almost no difference between the full sample and the one
containing only half the haloes. For even smaller subsamples, the
differences in the distributions are small, with no apparent systematic 
trend. Whereas the 25\% subsample appears to favour slightly smaller 
major branches, the 10\% subsample follows the opposite trend, with 
both deviating from the full sample distribution by small factors
only.

We thus conclude that while the downsampling affects the percolation
behaviour of the resulting halo samples, it does not change the 
basic properties of LSS as measured by our structure--finding
algorithm -- provided the comparison is done in such a way that the
largest object contains the same mass fraction in the different 
samples. This conclusion is supported by the earlier results obtained
by Bhavsar \& Splinter (1996).

\section{Summary and Discussion} \label{discussion}

We introduced a new algorithm to classify cosmic structures. This algorithm is
based on a Minimum Spanning Tree representation of groups of haloes that are
found with a standard FOF group finder. The new algorithm exactly reproduces 
the length distributions of filaments in a set of simple test cases.

Structures like the one shown in Figure~\ref{fig:slicemosaic} contain many thousands 
of structural elements, which in the context of this work were called major 
branches. We investigated the numbers and properties of these objects using 
haloes from the Millennium Run simulation (Springel et al. 2005). For the
group finding linking lengths between $b = 0.5$ and 0.6 were employed, around
the percolation threshold. For each linking length, we concentrate on the
largest object. We find that while the fraction of mass in the largest object 
and the actual number of its major branches are a function of $b$, the 
properties of those branches do not change\footnote{Note that the linking 
length, in physical units, in this range changes from 1.005\,$h^{-1}$\,Mpc to 
1.206\,$h^{-1}$\,Mpc or 2.01\% to 2.41\% of the size of the simulation 
volume in one dimension, while the fraction of halo mass in the largest object 
jumps from 2.1\% to 69.4\% -- the tell--tale sign of percolation.}.

We then computed quantities for the major branches that correspond to shapes 
and extents. Branches are predominantly planar, with straight or curved 
filamentary configurations preferred. A small number of branches appears to 
have significant extents in two dimensions. The shapes of the major branches 
are independent of the choice of $b$. This means that cosmological
percolation, while leading to a vast increase in the size of the object, 
does not alter its structural properties. Instead, as $b$ increases, more and 
more pieces of the same kind are being added to the largest object.

Large Scale Structure thus appears to be modular, with a fixed set of 
pieces -- with different sizes and shapes. This finding adds considerable
information to the earlier study by Colberg et al. (2005), who investigated
inter--cluster filaments. Note that here, we have not made any assumptions 
on where to look for filaments (or sheets). Given the difference in method, 
the general agreement between size distributions found here and in Colberg 
et al. (2005) is quite interesting. There, it was found that clusters with 
separations of up to 15\,$h^{-1}$\,Mpc almost always have a filament between 
them. Here, we find that the vast majority of branches have extents of up 
to 15\,$h^{-1}$\,Mpc.

It is reassuring that the results of this study support the visual impression
from images like the one shown in Figure~\ref{fig:slicemosaic}: LSS in N--body
simulations consists predominantly of a complex network of filamentary 
structures. This work represents the first systematic attempt to determine 
numbers, sizes and shape distributions of structure elements.

Different halo samples, chosen by mass, exhibit a correlation between the
minimum mass of the sample and the mean extent of the resulting major
branches. The larger the threshold mass, the longer the major branches.
Massive galaxy clusters form the backbone of LSS. It is reassuring to see 
that despite the differences between the methods employed here and in 
Bharadwaj et al. (2004), the largest structure elements extend over many 
dozens of Megaparsecs. With the longest single coherent structure element 
in the $m > 10^{14}\,h^{-1}\mbox{M}_\odot$ sample being almost 
200\,$h^{-1}$\,Mpc in length, the simulation appears to be in good shape to
account for very large features of LSS such as, for example, the SDSS 
Great Wall (Vogeley et al. 2004). 

The properties of major branches in LSS appear to be quite unfazed by
a downsampling of the halo sample to as little as 10\% of the original
size. For this test, for each halo we picked a linking length which led
to the largest object containing about 65\% of the total sample mass. 

It is tempting to argue that the predominance of filaments and the small 
number of sheets/walls is caused by how we investigate structures. After all,
it is possible that there are sheets in the simulation volume that the
the algorithm breaks up into many straight and warped filaments. In 
principle, it is hard to see what is wrong with this argument. However, 
when we visually inspected the distribution of haloes and the largest object 
we were unable to find prominent sheets. However, it appeared that most 
structures that we would classify as sheets appeared to consist of a set 
of connected filaments with gaps in between them. What this means is that 
while there are regions of space that do contain large mass concentrations 
in a planar configuration, inside that plane the matter has already collapsed 
into individual filaments. This not only supports the results obtained here, but
it also explains why studies using Minkowski Functionals and Shapefinders
(see Sheth \& Sahni 2005 and references therein) have not been able to find 
strong signals from sheets. 

Finding an algorithm to classify LSS would be a mostly academic exercise
if it was only applied to real--space haloes from Dark--Matter only
simulations. This current work merely introduces the algorithm, along
with some systematic tests. In a future study, we plan to apply the methods 
outlined here redshift--space data from mock galaxy catalogues and 
actual observed galaxy surveys.

\section*{Acknowledgments}

The Millennium Run simulation used in this paper was carried out by the
Virgo Consortium (http://www.virgo.dur.ac.uk) at the Computing Center of the 
Max--Planck--Gesellschaft in Garching, Germany. We thank Rupert Croft, 
Carlos Frenk, Tiziana di Matteo, Cameron McBride, Volker Springel, and 
Naoki Yoshida for helpful discussions and comments on earlier drafts of this 
work, and the anonymous referee for very helpful suggestions for
improvements. Thanks are also due to Esther Jesurum for advice concerning 
advanced C++ techniques and to Rien van de Weygaert for providing his code 
used in to prepare an earlier version of the structure--finding code. 

\label{lastpage}

\appendix

\section{Locating Tree Elements}

In a nutshell, structure elements of the MST are found by grouping its
individual elements into larger units. These larger units are constructed
on the basis of how the elements of the MST are interconnected. Before 
discussing the algorithm in detail, it is important that the notation
used in the folowing is clear. Our starting configuration is a MST,
constructed from cells on a grid. A {\it node} is simply such a cell,
that is an individual member of the MST. An {\it edge} is a connection 
between two nodes. Each node is connected to at least one other node,
and we will call the set of nodes that any given node is connected to its 
{\it neighbouring node(s)}. As already mentioned, the algorithm groups
nodes into larger units. Any such unit we will call a {\it branch}, and
it might contain a collection of connected nodes and other branches. 
Each branch {\it has to} contain at least one node. If a branch contains 
other branches then the longest of those branches will be called the 
{\it main branch}, with the remaining one(s) being {\it subbranches}. 
A branch is represented by the last node added to it.

\begin{figure*}
\includegraphics[width=105mm]{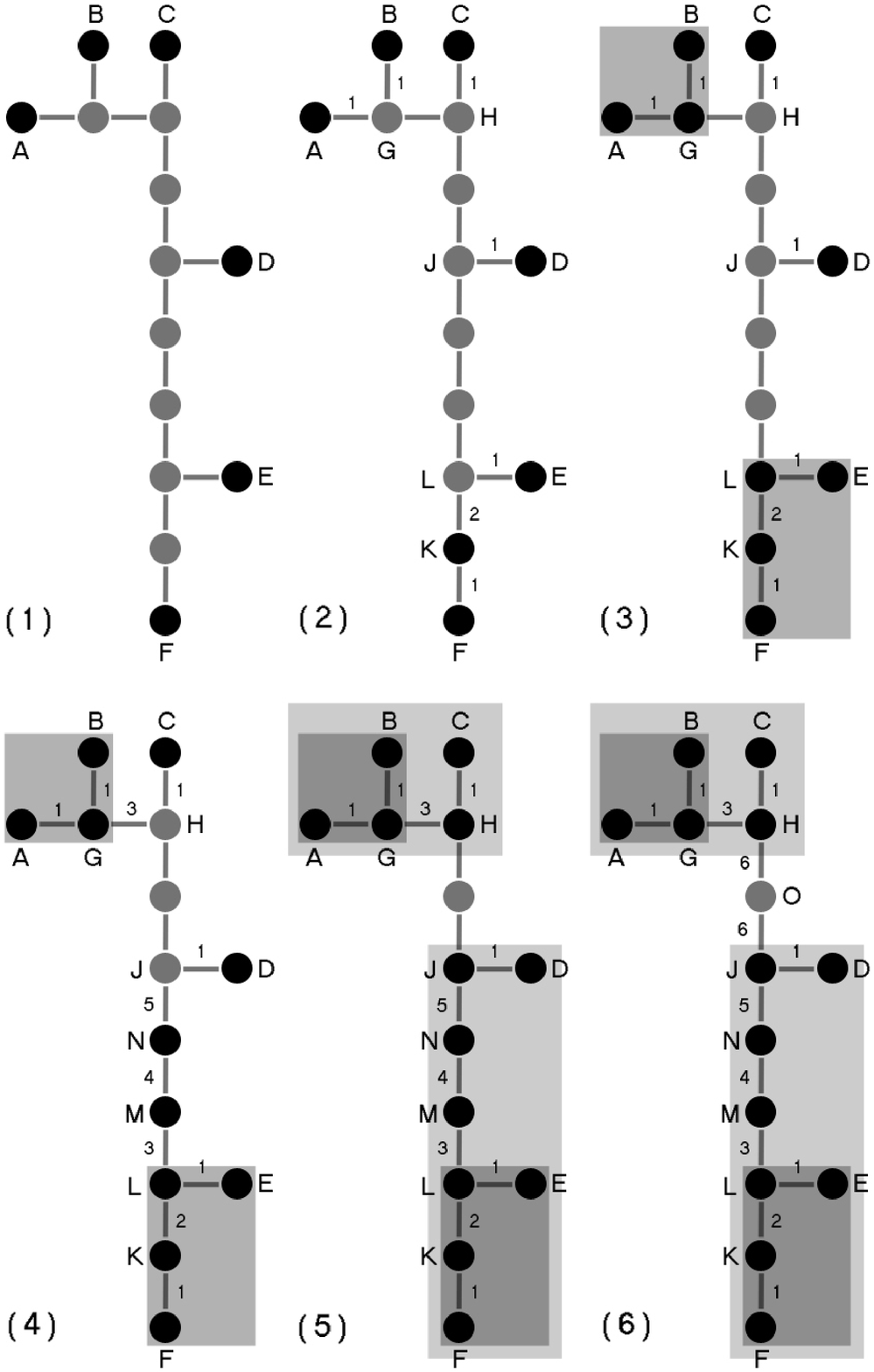}
\caption{A schematic overview of the algorithm to detect branches in the Minimal 
         Spanning Tree. Please refer to the main text for details on the 
         individual steps.} 
\label{tree_finding}
\end{figure*}

It is probably easiest to understant the classification algorithm by
referring to Figure~\ref{tree_finding}, which schematically depicts the
application of the algorithm to a very simple case. Figure~\ref{tree_finding}
shows the same MST at six different steps of the algorithm. As the
individual nodes are being grouped into units, the different branches
are shown as grey boxes around the nodes and edges they consist of.

The algorithm constructs branches from the MST's nodes and edges using a
simple set of rules. The rules are applied sequentially until each node is
contained in at least one branch.

The initial set of branches is constructed from all nodes that are connected
to only one other node -- the loose ends of the structure. In 
Figure~\ref{tree_finding}, these are nodes A to F. In order not to make
Figure~\ref{tree_finding} too cluttered, there are no grey boxes drawn
around these initial branches.

Using the set of initial branches, the full classification of the MST
is done by applying the following rules:
\begin{enumerate}
   \item As mentioned before, a branch is represented by the node that was
         added last. Take all branches that are not contained inside another
         branch and take their representing nodes. For those nodes, find 
         their neighbouring nodes. If a neighbouring node can only be added 
         to one node's branch (and to no other branch) add it to that branch, 
         and make it the new representing node of the branch. Increase the
         length of the branch accordingly.
   \item If a neighbouring node can be added to more than one branch, set 
         the node and the branch aside in a queue, and ignore this branch
         for the time being.
   \item For each branch, add only one neighbouring node at a time.
   \item Go back to the first step if (and only if) there are other nodes
         that can stil be added to a branch.
   \item If no node can be added to any branch, process the branches and nodes 
         in the queue. At this stage it is important to realize that each node
         that does not sit at the edge of the MST can be connected to a
         number $n$ of other nodes, and by construction of the algorithm,
         $n \ge 2$. What this means is that a node can potentially
         reside at the intersection of $n$ branches. In this step, the goal
         is to process all nodes for which there are either $n-1$ or $n$
         neighbouring branches contained in the queue, with the latter case
         only possible at the very end of the algorithm. Nodes for which
         less than $n-1$ neighbouring branches are in the queue are ignored.

         In the case where there are $n-1$ branches in the queue for a node
         consolidate these branches and the node as follows. Create a new
         branch, represented by the node, and add the $n-1$ branches as
         subbranches to the new branch. Find the longest subbranch from those
         and label it as the main subbranch. The length of the new branch
         is taken as the length of the main subbranch.

         As already indicated, the $n$ case will only be met once, at the very
         end of the algorithm (if this point is not clear, it will become 
         clear below, where we will discuss the simple example from 
         Figure~\ref{tree_finding}). The node and the branches in the queue
         are processed just like in the $n-1$, with the only differences
         being that a) the length of the final branch is taken as the sum of
         the two longest subbranches (there are at least two subbranches),
         and b) there is no more work left to do, so the algorithm is finished.
   \item After the queue has been cleaned up as far as possible continue
         adding the remaining nodes (if there are any left) by going back to
         the first step.
\end{enumerate}

It is probably worth discussing how the algorithm is applied in the case of
the simple MST depicted in Figure~\ref{tree_finding}. Please keep in mind,
though, that real cases will be threedimensional, and edges will not all
have the same length.

As already mentioned, step (1) in Figure~\ref{tree_finding} shows the 
identification of the initial set of branches by locating the loose
ends of the structures. These are the nodes labeled A to F.

Step (2) shows the application of rules (i) to (iv). For example, node
F is connected to node K, which at this stage can only be added to F
and to no other node. K is thus added to the branch and made the new
representing node. At nodes G, H, J, and L, rule (ii) applies. After the
first iteration, there is only one free node left to process, node K.
Its neighbour, L, sits at an intersection, so rule (ii) applies. After
this step, all nodes that have not been added to a branch (G, H, J, and
L) are inside the queue, and no free branch can be added anywhere.

Step (3) shows the application of rule (v). Figure~\ref{tree_finding} shows
that rule (v) describes how to deal with intersections, where several
branches meet at a node. In step (3), shaded boxes indicate which parts of
the queue can be processed. The case of node J is quite obvious. Node H
cannot be processed, yet, since node G itself is inside the queue. Nodes
G and L can be processed. At node G, two branches of the same size meet,
and one of them is picked as the main subbranch randomly. The new branch
contains the nodes A, B, and G, and two subbranches, namely A--G and B--G.
It is represented by node G. At node L, two branches with different sizes
meet, with F--K--L being longer than E--L. The new branch also has two
subbbranches, with the main subbranch given by F--K--L, and the new branch 
is represented by node L.

At step (4), we are back at applying rules (i) to (iv). At H, rule (ii)
applies immediately. From node L onwards, we can make our way towards 
node J (rules (i), (iii), and (iV)), at which point rule (ii) applies.

The remaining steps contain only one slightly difference, namely at step
(6), when processing the queue at node O, we are done. The final branch
contains all nodes and the subbranches constructed earlier, and the length
of the final branch is given by the length of the chain of nodes starting
at A, going to H and then down to F. We will refer to node O as the {\it 
base node}.

As Figure~\ref{tree_finding} shows, the classification of the MST is done
outside--in, that is the classification of the nodes into branches is done
such that larger branches are built from smaller ones. 

Once the MST has been categorized into sets of branches and subbranches, we 
construct another, somewhat simpler data structure from it. Starting at the 
base node {\it major branches} of the structure as follows: We build a new 
set of branches by now moving away from the base node and by adding nodes to 
these branches. Whenever we run into a node where the tree bifurcates, we 
ignore those subbranches that are shorter than some length scale $l$. The 
value of $l$ is an arbitrary pruning parameter (in the fashion of the pruning 
parameter used for MSTs discussed in Barrow et al. 1985). If there are two 
(or more) subbranches at some node which are longer than $l$, we create two 
new branches, add those to the current branch, and then we walk down those 
two (or more) branches. It needs to be said that this step really is just 
amounts to a rearrangement and slight pruning of the first classification.

Figure~\ref{major_branches} shows an example of a simple tree, with its major
branches marked (nodes that belong to a major branch are shown in black) and 
labeled (A to G). In this example, a subbranch is required to have length
2. Otherwise, its nodes will not be contained in any of the major branches.
Note that the major branches, unlike the branches in the original tree 
constructed from the MST, only contain nodes/edges. The major branches are
designed to represent the major structural elements in the tree.


\begin{thebibliography}{99}
\bibitem{babul92}        Babul A., Starkman G.D., 1992, ApJ, 401, 28
\bibitem{barrow85}       Barrow J.D., Bhavsar S.P., Sonoda D.H., 1985, MNRAS, 216, 17
\bibitem{barrow87}       Barrow J.D., Bhavsar S.P., 1987, QRAS, 28, 109
\bibitem{bharadwaj04}    Bharadwaj S., Bhavsar S.P., Sheth J.V., 2004, ApJ, 606, 25
\bibitem{bhavsar83}      Bhavsar S.P., Barrow J.D., 1983, MNRAS, 205, 61
\bibitem{bhavsar88a}     Bhavsar S.P., Ling E.N., 1988a, ApJ, 331, 63
\bibitem{bhavsar88b}     Bhavsar S.P., Ling E.N., 1988b, PASP, 100, 1314
\bibitem{bhavsar96}      Bhavsar S.P., Splinter R.J., 1996, MNRAS, 282, 1461
\bibitem{colberg05}      Colberg J.M., Krughoff K.S., Connolly A.J., 2005, MNRAS, 359, 272
\bibitem{colless01}      Colless M. et al., 2001, MNRAS, 328, 1039
\bibitem{croton05}       Croton D.J. {\it et al.}, 2005, astro--ph/0508046
\bibitem{david85}        Davis M., Efstathiou G., Frenk C.S., White S.D.M., 1985, ApJ, 292, 371D
\bibitem{dekel85}        Dekel A., West M.J., 1985, ApJ, 288, 411
\bibitem{demianski04}    Demianski M., Doroshkevich A.G., 2004, A\&A, 422, 423
\bibitem{doroshkevich01} Doroshkevich A.G., Tucker D.L., Fong R., Turchaninov V., Lin H., 2001, MNRAS, 322, 369
\bibitem{einasto83}      Einasto J., Klypin A., Shandarin S., 1983, IAUS, 104, 265
\bibitem{gott86}         Gott J.R., Dickinson M., Melott A.L., 1986, ApJ, 306, 341
\bibitem{gott05}         Gott J.R., Juri M. Schlegel D., Hoyle F., Vogeley M.,
                         Tegmark M., Bahcall N., Brinkmann J., 2005, ApJ, 624, 463
\bibitem{hopkins05}      Hopkins P.F., Bahcall N.A., Bode P., 2005, ApJ, 618, 1
\bibitem{klypin93}       Klypin A., Shandarin S.F., 1993, ApJ, 413, 48
\bibitem{krzewina96}     Krzewina L.G., Saslaw W.C., 1996, MNRAS, 278, 869
\bibitem{luo95}          Luo S., Vishniac E., 1995, ApJ, 96, 429
\bibitem{luo96}          Luo S., Vishniac E., Martel H., 1996, ApJ, 468, 62
\bibitem{mecke94}        Mecke K.R., Buchert T., Wagner H., 1994, A\&A, 288, 697
\bibitem{peacock99}      Peacock J.A., 1999, ``Cosmological Physics'', Cambridge University Press, Cambridge, UK, ISBN
                         0--521--42270--1
\bibitem{peebles75}      Peebles P.J.E., Groth E.J., 1975, ApJ, 196, 1
\bibitem{peebles}        Peebles P.J.E., ``The Large--Scale Structure of the Universe'', Princeton University
                         Press, Princeton, NJ
\bibitem{sahni98}        Sahni V., Sathyaprakash B.S., Shandarin S.F., 1998, ApJL, 495, 5
\bibitem{shandarin83}    Shandarin S.F., 1983, PAZh, 9, 195
\bibitem{shen05}         Shen J., Abel T., Mo H., Sheth R.K., astro--ph/0511365
\bibitem{sheth05}        Sheth J.V., Sahni V., 2005, ``Exploring the Geometry, Topology and Morphology of Large Scale
                         Structure using Minkowski Functionals'', to appear in Current
                         Science, also astro-ph/0502105
\bibitem{springel05}     Springel V., et al. (Virgo Consortium), 2005, Nature, 435, 629
\bibitem{springel01}     Springel V., White S.D.M., Tormen G., Kauffmann G., 2001, MNRAS,  328, 726
\bibitem{springel98}     Springel V. et al., 1998, MNRAS, 298, 1169
\bibitem{tago05}         Tago E., Einasto J., Einasto M., Saar E., 2005, astro--ph/0501099
\bibitem{turner76}       Turner E.L., Gott J.R., 1976, ApJS, 32, 409
\bibitem{vogeley04}      Vogeley M.S., Hoyle F., Rojas R.R., Goldberg D.M., ``Mapping the cosmic web 
                         with the Sloan Digital Sky Survey'', in: ``Outskirts of Galaxy Clusters: Intense 
                         Life in the Suburbs'', proceedings of the IAU Colloquium 195, ed. Diaferio A.,
                         Cambridge University Press, Cambridge, UK, ISBN 0--521--84908--X 
\bibitem{york00}         York D.G. et al., 2000, AJ, 120, 1579
\end{thebibliography}
\end{document}